\documentclass[aps,prb,reprint]{revtex4-2}

\draft 
\usepackage{amssymb,amsmath}
\usepackage{graphicx}
\usepackage{dcolumn}
\usepackage{bm}
\usepackage{color}
\usepackage{float}
\usepackage{hyperref}
\usepackage[version=4]{mhchem}
\usepackage{relsize}
\usepackage{comment}
\usepackage{media9}

\newcommand{\M}{{\cal M}}
\newcommand{\W}{{\cal W}}
\newcommand{\R}{{\cal R}}

\newcommand{\K}{{\cal K}}
\newcommand{\A}{{\cal A}}

\begin{document}

\title{Musical creativity enabled by nonlinear oscillations of a bubble in water}

\author{Ivan S.~Maksymov}
\email{imaksymov@csu.edu.au}
\affiliation{Artificial Intelligence and Cyber Futures Institute, Charles Sturt University, Bathurst, NSW 2795, Australia\looseness=-1}


\begin{abstract}
Producing original and arranging existing musical outcomes is an art that takes years of learning and practice to master. Yet, despite the constant advances in the field of AI-powered musical creativity, production of quality musical outcomes remains a prerogative of the humans. Here we demonstrate that a single bubble in water can be used to produce creative musical outcomes, when it nonlinearly oscillates under an acoustic pressure signal that encodes a piece of classical music. The audio signal of the response of the bubble resembles an electric guitar version of the original composition. We suggest, and provide plausible theoretical supporting arguments, that this property of the bubble can be used to create physics-inspired AI systems capable of simulating human creativity in arrangement and composition of music.
\end{abstract}

\maketitle 

\section{Introduction}
Bubbles in liquids underpin many important natural phenomena \cite{Bre95, Lau10, Mak22_comb, Mak22_Bio}, including cavitation \cite{Bre95} and sound of running water \cite{Min33, Lei87}. Oscillations of bubbles driven by an acoustic pressure wave are also similar to the behaviour of a biological brain since both the brain \cite{McK94} and the bubble \cite{Lau10} are nonlinear dynamical systems \cite{Str19}. Subsequently, studies of oscillating bubbles may help understand certain brain functions that are responsible for perception of sounds and music. The following experimental evidence speaks in favour of this proposition.

Firstly, it has been demonstrated that the electric charge impulses that underpin the nerve signalling are accompanied by acousto-mechanical (sound) waves that are intrinsically nonlinear \cite{Hei05, Gon14, Had15, Mak20}. Secondly, it has been shown that human experience in musics is mediated by nonlinear-acoustical processes \cite{Mak19} and that the same processes underpin the auditory processing abilities of some animals \cite{Lev06}. In particular, in an experiment involving owls exposed to a piece of classical music made up of tones with deliberately removed fundamental frequency harmonics, the owl's brain restored the missing fundamental harmonics \cite{Jan96}. While such a behaviour is of considerable interest in the field of nonlinear physics \cite{Mak19_Faraday}, effectively the owl's brain transferred a musical idea from its original position to a lower frequency, which is a common examples of octave transposition \cite{Lev06}. Significantly, while transposition in musics is a nontrivial task that is accessible mostly to individuals with a formal relevant education, a biological brain can do this type of audio processing naturally. Thirdly, it is also well-known that nonlinear-acoustical processes \cite{Mak19} underpin the operation of many musical instruments \cite{Fle90} and that musicians, as well as many people who love music but do not have formal background in it, understand nonlinear effects naturally without knowing much about nonlinear physics \cite{Lev06}. This fact also indicated that a biological brain can naturally process nonlinear acoustic signals.  

Recently, we suggested that a cluster of oscillating bubbles in water can operate as an artificial neural network that exhibits complex nonlinear behaviour and that can be trained to predict highly nonlinear and chaotic time series that arise in many practical situations \cite{Mak21_ESN} such as the analysis of financial markets, weather forecasting and control of autonomous vehicles \cite{Luk09, Nak20, Tan19, Nak21}. Thus, since the particular kind of the artificial neural networks that oscillating bubbles can efficiently emulate---the Echo State Network (ESN) \cite{Luk09} and Liquid State Machine (LSM) \cite{Maa02}---can also reproduce some functions of a biological brain, it is conceivable that oscillating bubbles may also reproduce some of the brain's functions, including those associated with the perception of music. Hence, in this work we suggest that a highly nonlinear behaviour of an oscillating bubble could be used to complete some tasks that require musical creativity.

Musical creativity can be defined as a process of employing existing musical knowledge to produce novel musical outcomes that may take, for example, the form of improvisations, compositions and arrangements. The production of quality musical outcomes remains one of the most challenging tasks for machine learning systems \cite{Mil19} despite the recent significant progress in AI-powered musical creativity (see \cite{Pac12, Hua16, Bri21, Her23} to cite a few works). Subsequently, the idea that a simple bubble could produce a creative musical output is not only fundamentally intriguing but can also lead to new knowledge in the field of AI.

As a representative example, we synthesise a simple version of ``In the Hall of the Mountain King'', a piece of music composed by Edvard Grieg, and use it as the acoustic signal that drives nonlinear oscillations of a single bubble in water. By means of rigorous numerical simulations, we demonstrate that the output signal produced by the bubble is perceived as a ``heavy metal'' cover on the original composition decorated with warm and gritty tones typical of an electric guitar \cite{Ross}.
\begin{figure*}[t]
  \centerline{\includegraphics[width=0.9\textwidth]{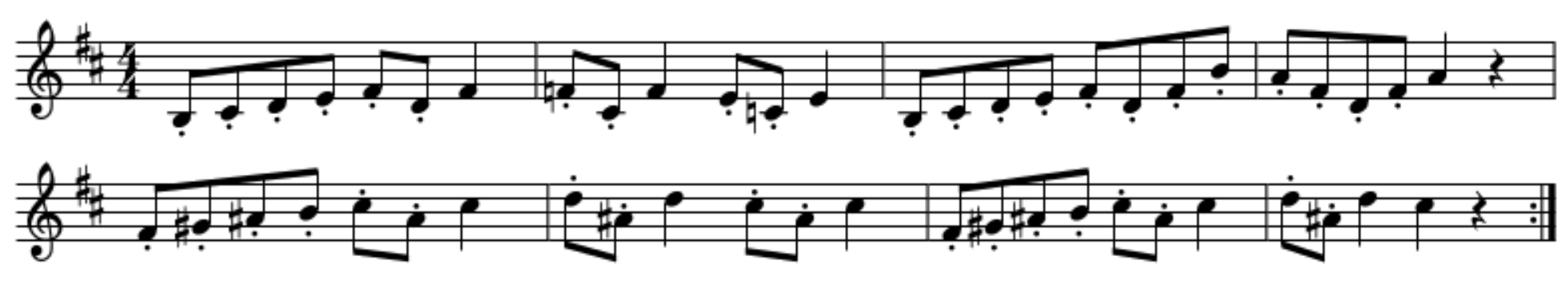}}
  \caption{Notes of the piece of music used as the acoustic signal that drives the oscillations of the bubble.\label{fig:fig1_notes}}
\end{figure*}

This paper is organised as follows. Our main findings are presented in Sect.~\ref{sec:Results} and followed by the discussion in Sect.~\ref{sec:Discussion}. The discussion is supported by a comprehensive theoretical analysis and numerical modelling of the physical properties of an acoustically driven oscillating bubble, the results of which are presented in Sect.~\ref{sec:Num} and Sect.~\ref{sec:Square}. Since memory is one of the prerequisites of creativity \cite{Ben23, Ger22}, we employ the ESN algorithm described in Sect.~\ref{sec:ESN} to demonstrate in Sect.~\ref{sec:Memory} that an oscillating bubbles possesses a memory capacity suitable for applications in the field of AI.

\section{Results\label{sec:Results}}
We choose a simple piano version of ``In the Hall of the Mountain King'' by Edvard Grieg to be the acoustic signal driving oscillations of the bubble (Fig.~\ref{fig:fig1_notes}). This composition is well-known to the general public and is also often used by musicians to produce their own recordings. Relying on the principles of 8-bit computer music arrangement \cite{Wau85}, we encode each note of the melody as a sequence of square pulses repeated at the frequency of the $n$th key of an idealised acoustic piano:
\begin{equation}
  \label{eq:eq12}
  f(n) = 2^{\frac{n-49}{12}} \times 440\,\text{Hz}\,.
\end{equation}
A four beats per bar time signature of the melody and its 120\,bpm tempo also enable us to calculate the duration of each bar in seconds. Although this approach cannot be used reproduce the exact sound of a piano, it suffices to create an easily recognisable version of the composition (see the supplementary audio file \texttt{out.in}) in a format suitable for processing by the numerical model employed in this work.

In the numerical model, we consider a single mm-sized bubble. The fundamental nonlinear physics underlying the interaction of a mm-sized bubble with a single square acoustic pressure pulse is discussed in Sect.~\ref{sec:Square}. Here we analyse the response of the bubble to a sequence of identical square pulses that encode the melody. We also record the response of the bubble as an audio file (see the supplementary file \texttt{out.wav}) and we discuss the aesthetic characteristics of the resulting audio record.

While the results presented in this section were obtained using positive square pressure pulses that increase the ambient pressure of the bubble in a stepwise manner, similar physical behaviour of the bubble was also observed using negative square pulses. Pulses of different peak amplitude and temporal duration were also considered. We established that the choice of the type of the pulses and of their particular duration and peak amplitude influences the aesthetic characteristics of the output produced by the bubble. Therefore, instead of being guided by the physics of interaction between the bubble and acoustic pressure pulses, our choice of the model parameters is dictated by the goal of achieving an appealing aesthetic characteristic.

\begin{figure}[t]
  \centerline{\includegraphics[width=0.5\textwidth]{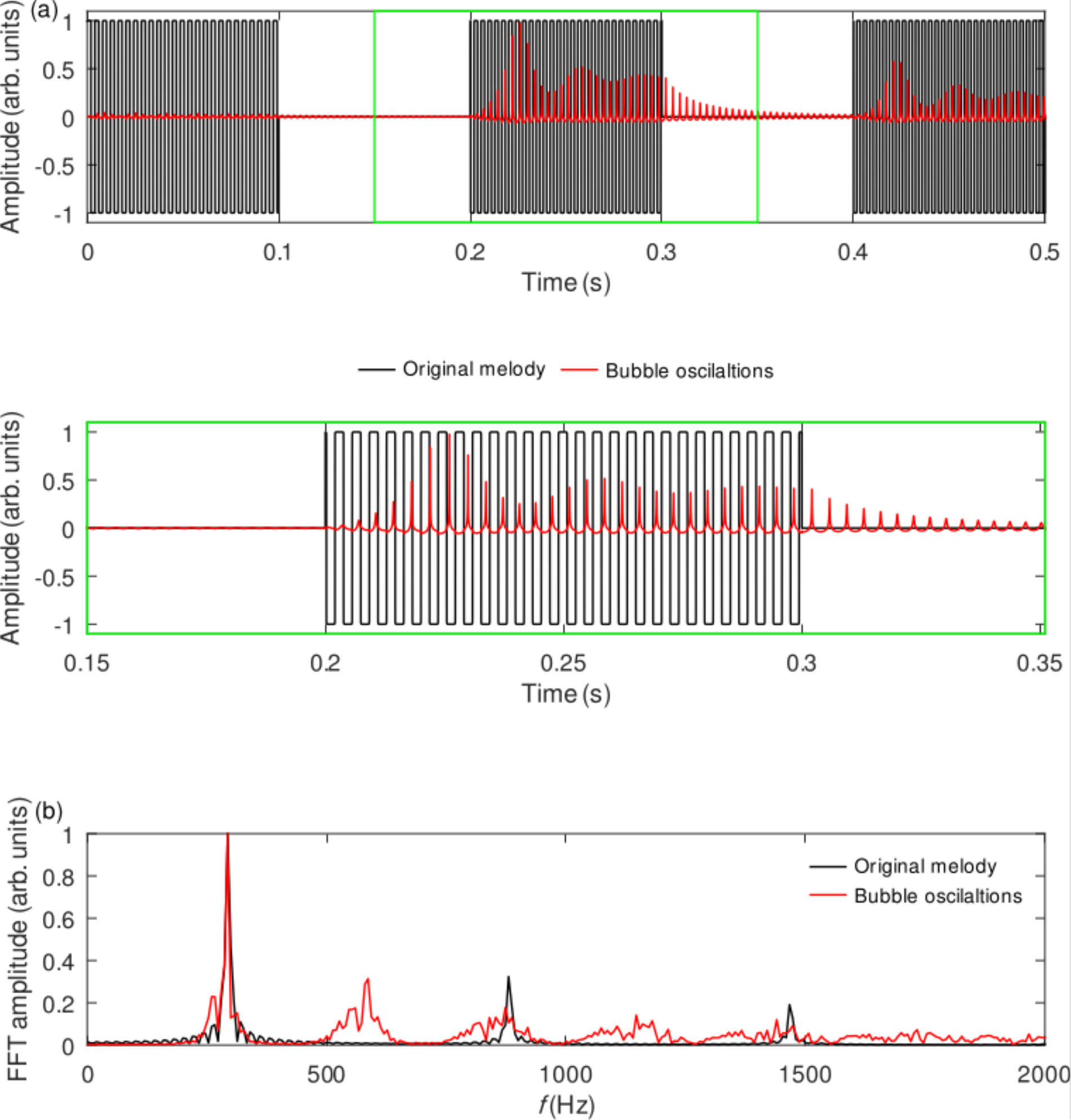}}
  \caption{(a)~Waveforms corresponding to three notes of the input melody (the black curve) and the respective acoustic response of the bubble (the red curve). The green rectangle highlights the time range of a magnified portion of the signals plotted in the inset located below the main panel. (b)~Fourier power spectra of the waveforms shown in the inset to panel~(a).\label{fig:fig3}}
\end{figure}

In Fig.~\ref{fig:fig3}(a) we plot several input square pulses corresponding to the notes of the melody (the black curve) and compare them with the respective acoustic response of the bubble (the red curve). Nonlinear response of oscillating bubbles has been the subject of many theoretical and experimental works (see \cite{Pro74, Bre95, Lau10, Sus12, Mak21} to cite a few), and we establish that the response of the bubble to the acoustic signature of the melody is also highly nonlinear. Moreover, we can see that the bubble continues oscillating during the periods of time between the individual pulses associated with the musical notes. This result speaks in favour of the ability of the oscillating bubble to have memory, which is an essential property of a nonlinear processing unit suitable for application in ESN as well as a prerequisite for artificial creativity (see Sect.~\ref{sec:Memory}).

In particular, we argue that a prolonged oscillation of the bubble excited by the sound of the composition is equivalent to the sustain effect used in the music instruments such as electric guitar and piano, where the length of time a note audibly resonates is deliberately prolonged \cite{Ross}. This behaviour can also correspond to the so-called nonlinearity with memory employed in advanced digital implementations of electric guitar distortion effects \cite{Yeh08, Yeh_thesis}. There, an idealised nonlinear system with memory can be represented analytically as a Volterra series, where the output of the nonlinear system depends on the input to the system at all other times, thereby providing the ability to create fading memory \cite{Boy85}.   

Indeed, as shown in Fig.~\ref{fig:fig3}(b), in the frequency domain the nonlinearity of the bubble manifests itself as the enrichment of the spectrum with higher-order harmonics---compare the black curve corresponding to the spectrum of the acoustic signature of the melody with the red curve corresponding to the response of the bubble. In electric guitar performances, the appearance of higher-order harmonics is associated with fuzzy and gritty tones \cite{Ross, Zol, Yeh08}. Subsequently, the recorded output signal (see the supplementary file \texttt{out.wav}) is aesthetically perceived as an electric guitar cover of the original piece of music. 

To test our perception, we asked people with formal music education to listen to the output signal and they confirmed that the melody indeed closely resembles an electric guitar version of the original melody. We also used Audacity audio processing software,  where we applied the standard static nonlinear audio processing effects to the original melody to simulate the effect of an electric guitar \cite{Zol, Yeh08}. The goal of this procedure was to reproduce the lineshape and spectrum of the bubble response as well as to recreate the electric guitar effect of the bubble on the original composition.  

\begin{figure}[t]
  \centerline{\includegraphics[width=0.5\textwidth]{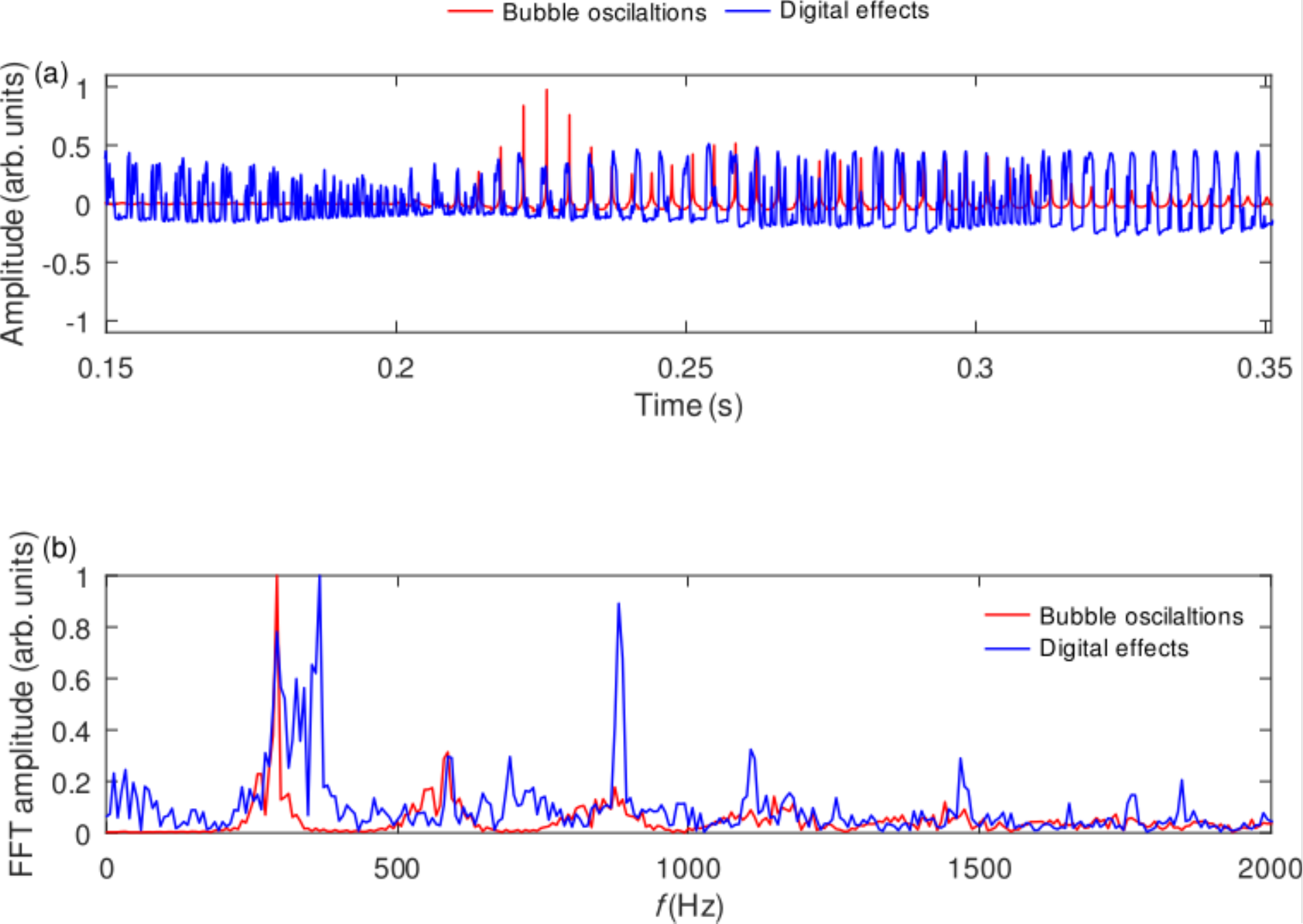}}
  \caption{(a)~Comparison of the temporal response of the bubble (the red curve) with a signal obtained from the original melody by digitally applying reverberation and nonlinear acoustic effects (the blue curve). (b)~Fourier spectra of the signals in panel~(a). The same times scale as in Fig.~\ref{fig:fig3}(a) is used. The waveform of the original melody is not shown for the sake of clarity.\label{fig:fig3_1}}
\end{figure}

The result of application of the digital effects is shown in Fig.~\ref{fig:fig3_1}, where we applied reverberation, which is a process relevant to both memory in artificial intelligence \cite{Kir91} and to cognitive memory \cite{Rib04}, and a guitar distortion effect that enriches the spectrum with the higher-order harmonics of the fundamental frequency \cite{Zol, Ross} (the nonlinearity of this distortion effect is memoryless \cite{Yeh08, Yeh_thesis}, which justifies the addition of the reverberation effect to our model). The lineshape and the spectral composition of the digitally produced signal (the blue curve) resemble those of the response of the bubble (the red curve). Since the use of distortion is characteristic of the heavy metal style in music \cite{Zol}, the fact that the bubbles reproduces this effect serves as an objective confirmation of its ability to creatively process music.  

\section{Discussion\label{sec:Discussion}}
Thus, we show that the ability of a single oscillating bubble to perform complex nonlinear tasks enables it to arrange existing music pieces similarly to a human. Creative music composition using the properties of liquids is an established style of music \cite{sound_art}, where it is highly likely that nonlinear acoustic effects associated with bubbles trapped in liquids \cite{Min33, Lei87} have already been exploited in some form.

However, we look at the nonlinear properties of the bubble at a different angle. In our previous work we demonstrated the ability of oscillating bubbles to forecast chaotic time series similarly to an artificial neural network \cite{Mak21_ESN}, which is a task that requires not only nonlinearity but also memory. Most importantly, unlike a bubble in water that produces some sounds that are then used by an artist to compose music, a bubble employed in an artificial neural network plays an active role of an analog data processing unit that mimics the operation of a biological neuron. Although the memory capacity and the speed of data processing of such a unit are low compared with those of a typical digital computer, it has been demonstrated that analog computer systems can be more efficient than digital ones in solving certain classes of problems pertinent to the field of AI \cite{Nak21, Mak21_ESN, Cao22, Iva22}. Given this, we suggest that a single bubble can be employed as a building block of an analog AI system that can produce musical outputs with no or little human input. 

Unlike the music transcription, which is an exact note for note rendition of a piece of music written for one instrument and played on another (i.e.~piano to guitar), arrangement is a more creative process, where the style of the music is changed and new complex tones are added. Consequently, this creative task is especially challenging for AI system because it requires a machine to have some of the key features of the human intelligence such as the ability to associate ideas, perceive, think, search for answers and criticise results of own work \cite{Bod98}. Yet, creativity relies on cognitive memory \cite{Ben23, Ger22} and is closely linked to cultural context and personality, also being influenced by motivation and emotions of the artist \cite{Bod98}. Interestingly enough, the ability to appreciate, arrange and compose heavy metal music has also been associated with high intellectual abilities \cite{Cad07}, which means that the production of heavy metal style music should be a particularly challenging task for AI.

Thinking in terms of machine learning systems, researchers have attempted to create models, where training inputs for achieving artificial creativity are represented by poorly defined data sets affected by perturbations and noise \cite{Tha16, Ole19}. The studies of so-created models have revealed that achieving artificial creativity may contradict the standard approach to training an artificial neural network since perturbations associated with creativity interfere with the operation of the network, for instance, by altering the values of its connection weights \cite{Tha16}. Subsequently, it has been suggested that the organisational principles of conventional neural network should be changed to enable AI-powered creativity \cite{Ole19}.

However, this assessment of the applicability of conventional artificial neural networks in the field of AI creativity does not take into account the recent advances in the development of analog (non-digital) counterparts of neural networks, where hardware and real-life physical systems that exhibit a nonlinear dynamical behaviour are used as artificial neurons. While such physical computation systems have thus far mimicked the operation of some digital neural network architectures \cite{Nak21}, it has been demonstrated that they hold a potential to surpass the abilities of a computer program in operations intended to simulate the functions of a biological brain \cite{Nak21, Cuc22, Mak21_ESN}. The findings presented in this work contribute to the endeavour to demonstrate this potential.      

Finally, assuming that an AI system has been able to creatively generate a musical output comparable with that produced by a human, researchers face yet another problem: the quality of the AI-generated musical output is difficult to assess since this process would rely on the appreciation of trained listeners \cite{Pac12}, who may, in turn, hold a cultural bias \cite{Dem16} or a bias against computer-composed music \cite{Pas16}.

Since it is challenging to critically judge the aesthetical quality of the electric guitar arrangement produced by a bubble, we found several covers on ``In the Hall of the Mountain King'' produced by professional electric guitar players \cite{King, King1}. Even though those compositions sound more appealing than the version produced by the bubble, we have been able to distinguish the same warm and gritty tones created by the professional musicians in their performances. Thus, we leave it up to the readers to listen to the cited compositions and to comparatively judge the quality of the output produced by the bubble. However, we note that while it takes years of practice for a human to master an electric guitar, a single bubble in water appears to have an intrinsic ability to reproduce the sound of this musical instrument.
\\
\section{Conclusions}
Using a rigorous numerical model of nonlinear oscillations of an acoustically-driven single bubble in water, we have demonstrated that a bubble can produce audio outputs that aesthetically sound as creative musical outcomes produced by humans. Since past research demonstrated that oscillating bubbles can form a physics-based artificial neural network that can simulate certain functions of a biological brain, we suggest that either a single oscillating bubble or a network (cluster) of such bubbles could be used as an apparatus capable of reproducing some forms of artificial creativity. Achieving creativity in musics has thus far been a challenging to resolve problem for modern AI systems. Therefore, we believe that our findings may contribute to further development in this vital field of fundamental and applied research, also being of interest to artists, who experiment with the acoustic properties of liquids. 

\begin{acknowledgments}
ISM thanks Professor Sergey Suslov and Dr Andrey Pototsky (Swinburne University of Technology) for valuable discussions, and Professor Mikhail Kostylev (The University of Western Australia) for help with the calculation of the memory capacity.
\end{acknowledgments}

\bibliography{V1_new}

\appendix

\section{Numerical model of bubble oscillations\label{sec:Num}} 
The accepted model of nonlinear oscillations of a single gas bubble driven by an acoustic pressure wave is given by the Keller-Miksis (KM) equation \cite{Kel80} that takes into account the decay of bubble oscillations due to viscous dissipation and fluid compressibility:
\begin{eqnarray} 
  \left(1-\frac{\dot{R}}{c}\right)R\ddot{R}
  +\frac{\dot{R}^2}{2}\left(3-\frac{\dot{R}}{c}\right) = \nonumber\\ 
  \frac{1}{\rho}\left[1+\frac{\dot{R}}{c}+\frac{R}{c}\frac{d}{dt}\right]
  \left[P(R,\dot{R})-P_\infty(t)\right]\,,\label{eq:eq1}
\end{eqnarray}
where overdots denote the differentiation with respect to time $t$ and
\begin{equation}
  P(R,\dot{R})=\left(P_0-P_v+\frac{2\sigma}{R_0}\right)
      \left(\frac{R_0}{R}\right)^{3\kappa}
  -\frac{4\mu\dot{R}}{R}-\frac{2\sigma}{R}\,.\label{eq:eq2}
\end{equation}
The expression $P_\infty(t)=P_0-P_v+\alpha P_a(t)$, where $P_a(t)$ is an arbitrary driving acoustic pressure signal, represents the time-dependent pressure in the liquid far from the bubble. Parameters $R_0$, $R(t)$, $\mu$, $\rho$, $\kappa$, $\sigma$, $c$ and $\alpha$ denote the equilibrium and instantaneous radii of the bubble, the dynamic viscosity and the density of the liquid, the polytropic exponent of a gas entrapped in the bubble, the surface tension of a gas-liquid interface, the speed of sound in the liquid, and the amplitude of a driving ultrasound wave. Diffusion of the gas through the bubble surface is neglected. 

When bubble oscillations are not affected by fluid compressibility, which is the case in this work, the acoustic power scattered by the bubble in the far-field zone is \cite{Bre95}
\begin{equation}
  \label{eq:eq3}
  P_{scat}(R,t) = \frac{\rho R}{h}\left(R\ddot{R}+2\dot{R}^2\right)\,, 
\end{equation}
where $h$ is much larger than the equilibrium radius of the bubble.
\begin{figure*}[t]
  \centerline{\includegraphics[width=0.95\textwidth]{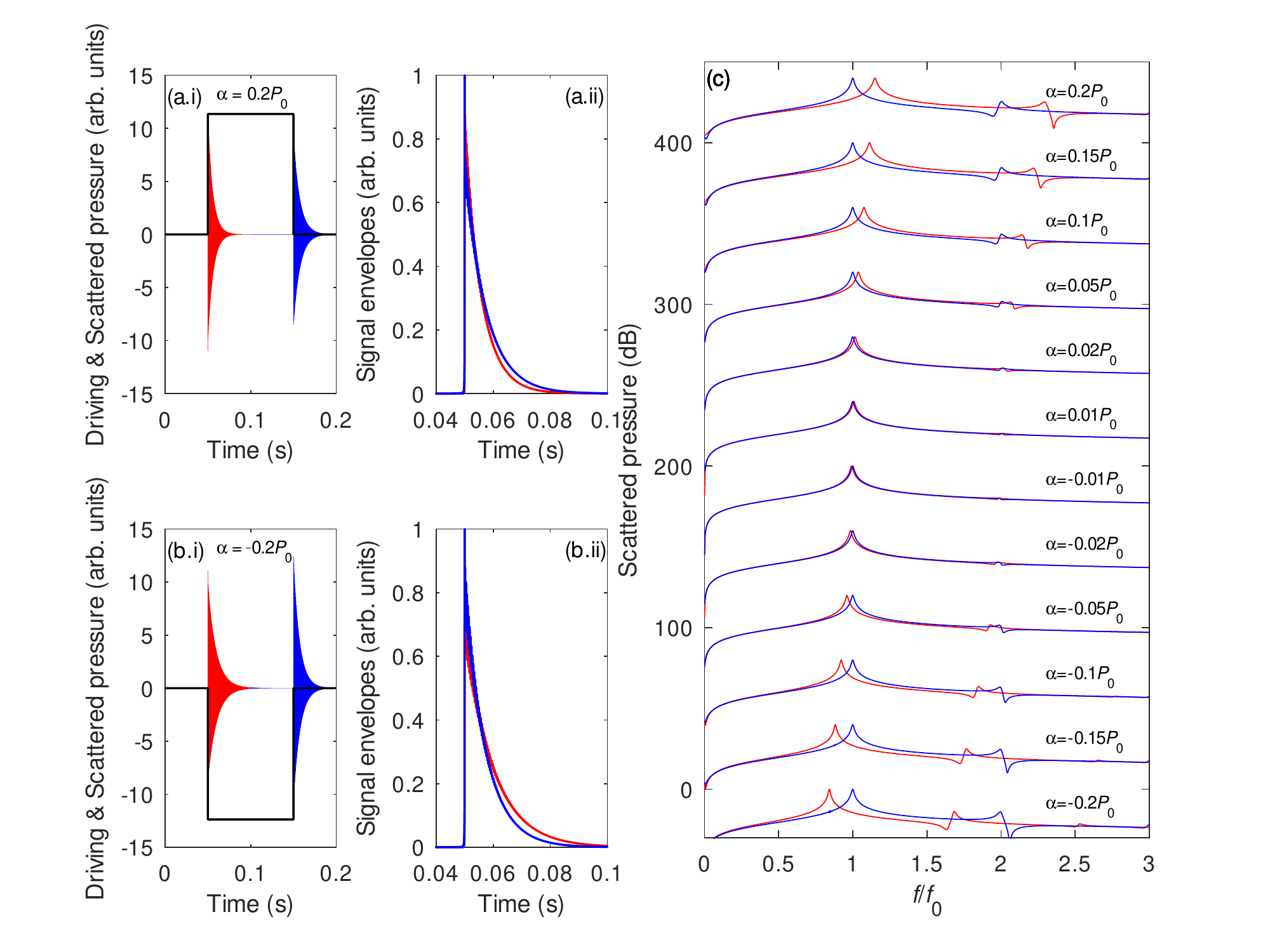}}
  \caption{Acoustic response of a single mm-sized bubble in water to a (a.i)~positive and (b.i)~negative pressure pulse. The respective, normalised to unity, envelopes of the bubble responses to the step changes in acoustic pressure are plotted in panels (a.ii) and (b.ii). (c)~Spectra of the pressure scattered by the oscillating bubble driven by the square pulses with different peak amplitudes $\alpha$. The red (blue) curves denote the spectra of the bubble oscillations caused by the first (second) step change in the pressure of the driving square pulse. As a result of the second step change, the bubble always returns to its equilibrium state and its fundamental oscillation frequency frequency remains unchanged.\label{fig:fig1_new}}
\end{figure*}

Since oscillations of the the bubble are excited by a signal consisting of square pulses (see Sect.~\ref{sec:Results}), we introduce a pulse repetition frequency parameter $f_p$ and rewrite Eqs.~(\ref{eq:eq1})--(\ref{eq:eq2}) in the nondimensional form \cite{Sus12, Mak21}, where we use the equilibrium radius of the bubble $R_{0}$ and ${\omega_p}^{-1}$ ($\omega_p = 2\pi f_p$) as the length and time scales, respectively, also introducing the non-dimensional quantities $r={R(t)}/R_{0}$ and $\tau = \omega_p t$. We obtain
\begin{equation}
\begin{gathered}
  {r}''[(1-\Omega r')r+\Omega\R] = (\Omega r'-3)\frac{{{r}'}^2}{2} 
  -\frac{\W+\R r'}{r} \\
  +(\M+\W)\frac{[1+(1-3\kappa)\Omega r']}{r^{3\kappa}} \\
  -(1+\Omega r')(\M+\M_e P_a(\tau))-\M_e \Omega r{P_a}'(\tau)\,,\label{eq:eq4}
\end{gathered}
\end{equation}
where primes denote the differentiation with respect to $\tau$,
$\Omega=\dfrac{\omega_p R_0}{c}$,
$\R=\dfrac{4\mu}{\rho\omega_p R^2_{0}}$,
$\W=\dfrac{2\sigma}{\rho{\omega_p}^{2}R^3_{0}}$,
$\M=\dfrac{P_0-P_v}{\rho{\omega_p}^{2}R^2_{0}}$,
$\M_e=\dfrac{\alpha}{\rho{\omega_p}^{2}R^2_{0}}$
and $\K=3\kappa$. Parameter $\Omega$ characterises the bubble size relative to the acoustic wavelength, $\M$ characterises elastic properties of the gas and its compressibility, $\W$ and $\R$ can be treated as inverse Weber and Reynolds numbers, representing the surface tension and viscous dissipation effects, respectively, and $\M_e$ is the measure of the acoustic forcing \cite{Sus12}. 

Computations were carried out for the following fluid parameters corresponding to water at $20^\circ$\,C: $c = 1484$\,m/s, $\mu=10^{-3}$\,kg\,m/s, $\sigma=7.25\times10^{-2}$\,N/m, $\rho=10^3$\,kg/m$^3$ and $P_v=2330$\,Pa. We take the air pressure in a stationary bubble to be $P_0=10^5$\,Pa and the polytropic exponent of air to be $\kappa=4/3$. In the calculations, the pulse repetition frequency $f_p$ depends on the notes of the piece of musics and it is always smaller than the frequency of natural oscillations $f_0$ of mm-sized bubbles ($f_0R_0\approx3.26$\,m s$^{-1}$
\cite{Lau10} and for $R_0=1$\,mm $f_0 = 3260$\,Hz). Equation~(\ref{eq:eq4}) is solved numerically using a fixed-step fourth order Runge-Kutta method implemented in a customised subroutine \texttt{rk4} \cite{Mudrov} ported from Pascal to Oberon-07 programming language.

We calculate the typical values of the nondimensional parameters $\M$, $\W$, $\R$ and $\Omega$ setting $R_0=1$\,mm,  $\alpha=0.2P_0$ and $f_p=100$\,Hz: 
\begin{equation}
  \R = \frac{\R_0}{\Omega^2}, \, \W = \frac{\W_0}{\Omega^3}, \, 
  \M = \frac{\M_0}{\Omega^2}, \, \M_e = \frac{{\M_e}_0}{\Omega^2}\,,
  \label{eq:eq5}
\end{equation}
where
\begin{eqnarray}
  \nonumber 
  \R_0 = \frac{4 \mu \omega_p}{\rho c^2} \approx 1.14\times10^{-9}\,  \\ 
  \W_0 = \frac{2 \sigma \omega_p}{\rho c^3} \approx 2.79\times10^{-11}\,  \\ \nonumber
  \M_0 = \frac{P_0-P_v}{\rho c^2} \approx 4.43\times10^{-5}\,  \\ \nonumber
  {\M_e}_0 = \frac{\alpha}{\rho c^2} \approx 9.08 \times10^{-5}\,, 
  \label{eq:eq6}
\end{eqnarray}
for $\Omega=4.54\times10^{-6}$. At these parameters the value of $\M$ exceeds those of $\W$ and $\R$ by several orders of magnitude, which represents the well-known fact that the dynamics of large gas bubbles is mostly determined by the gas elasticity, while both viscous dissipation and surface tension play secondary roles \cite{Sus12}.

\section{Bubble forced by a unit step change in acoustic pressure \label{sec:Square}}
The response of a single bubble to a unit step change in the driving acoustic pressure 
was previously investigated in the context of interaction between two oscillating bubbles \cite{Pel93}. Here, we investigate it in the context of sound processing since the piece of music used to excite oscillations of the bubble is encoded using a series of square pulses (see Sect.~\ref{sec:Results}).

We define the driving acoustic signal as $P_a(\tau)=\Theta(\tau)$, where $\Theta(\tau)$ is the Heaviside function
\begin{equation}
  \Theta(\tau) =
    \begin{cases}
      0 & \text{for }\,\tau < 0\\
      1 & \text{for }\,\tau > 0\,.
    \end{cases}       
\end{equation}
We assume that the nondimensional driving sound pressure amplitude $\M_e$ is small and that the bubble oscillates near its equilibrium state so that its instantaneous nondimensional radius is $r(\tau) = 1 + r_1(\tau)$, where $r_1(\tau) \ll 1$. We linearise Eq.~(\ref{eq:eq4}) around $r=1$ to obtain the following equation of a damped linear harmonic oscillator subjected to constant forcing 
\begin{equation}
  (\Omega\R+1){r_1}'' + (\R+\M_e\Omega+\A\Omega){r_1}'
     + \A r_1 = -\M_e\,,\label{eq:eq1_linear}
\end{equation}
where $\A \equiv (3\W+3\M)\kappa-\W$. Then we denote 
$Q = \dfrac{\Omega\R+1}{\R+\M_e\Omega+\A\Omega} \approx \dfrac{\Omega}{\K\M_0}$,
$\Lambda = \dfrac{\M_e}{\Omega\R+1} \approx \dfrac{\M_{e0}}{\Omega^2}$
and $\omega_0 = \sqrt{\dfrac{A}{\Omega\R+1}} \approx\sqrt{\K\M} =
\dfrac{\sqrt{\K\M_0}}{\Omega}$, where we used the fact that $\M \gg \W, \R$ established in Sec.~\ref{sec:Num}, and we recast Eq.~(\ref{eq:eq1_linear}) as an equation of an unforced harmonic oscillator 
\begin{equation}
  {r_1}'' + \frac{1}{Q}{r_1}' + 
    \omega_0^2(r_1+\frac{\Lambda}{\omega_0^2}) = 0\,,\label{eq:eq2_linear}
\end{equation}
where parameter $Q$ is the quality factor that describes the resonance behaviour of the harmonic oscillator and $\omega_0$ corresponds to undamped natural frequency of the bubble. With the initial conditions $r_1(0^-)={r_1}'(0^-)=0$ the solution of
Eq.~(\ref{eq:eq2_linear}) is
\begin{eqnarray} 
  && r_1 = e^{-\frac{\tau}{\tau_0}}
 \left(\frac{\Lambda\sin{\omega'\tau}}{2Q\omega_0^2\omega'} 
 + \frac{\Lambda\cos{\omega'\tau}}{\omega_0^2} \right)
 - \frac{\Lambda}{\omega_0^2}\,, \label{eq:eq3_linear}
\end{eqnarray}
where $\omega' = \sqrt{\omega_0^2-\frac{1}{4Q^2}}$ is the damped natural frequency of the bubble and $\tau_0=2Q \approx \dfrac{2\Omega}{\K\M_0}$
is the characteristic relaxation time over which the magnitude of oscillations in the transient regime reduces by the factor of  $e$.
 
Thus, apart from a transient response, the solution also contains a time-independent response $-\Lambda/\omega_0^2$. Therefore, one can see that under constant forcing the bubble oscillates near a changed equilibrium state $ 1+r_1(\tau\to\infty)=1-\frac{\Lambda}{\omega_0^2}$, which in the framework of the adopted model means that the equilibrium radius of the bubble has decreased. This is consistent with the fact that the constant forcing has effectively increased the static pressure outside the bubble. As a result, the oscillation frequency of the bubble is increased since $\omega_0 \sim \Omega^{-1}$ but the relaxation time is decreased since $\tau_0 \sim \Omega$ thus accounting for the fact that smaller bubbles have a higher stiffness \cite{Akh97} (recall that parameter $\Omega$ characterises the equilibrium radius of the bubble). 

This behaviour is confirmed by a direct numerical solution of Eq.~(\ref{eq:eq4}) for a single square pulse forcing term $P_a(\tau)$ defined as 
\begin{equation}
  P_a(\tau) =
    \begin{cases}
      A & \text{for }\,\dfrac{-\tau_p}{2} \leq \tau \leq \dfrac{\tau_p}{2}\\
      0 & \text{otherwise}\,,
    \end{cases}       
\end{equation}
where the duration of the pulse is $\tau_p>\tau_0$, the unit amplitude of the pulse $A$ can be either $-1$ or 1 so that the effects of bubble forcing with both positive and negative square pulses can be accounted for, and the strength of the forcing is defined by parameter $\M_e$ that is a function of the dimensional pressure amplitude $\alpha$.
\begin{figure}[t]
  \centerline{\includegraphics[width=0.4\textwidth]{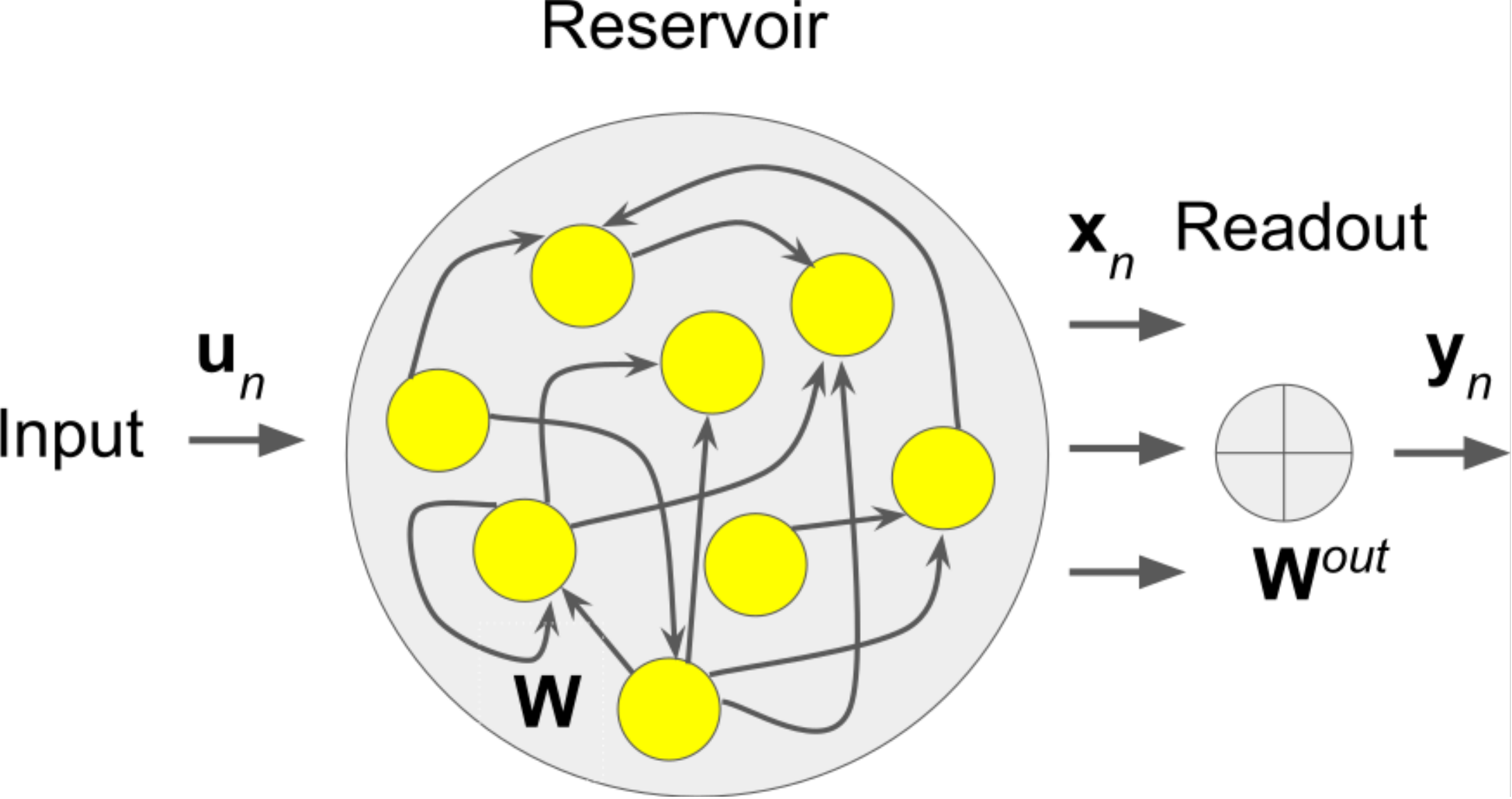}}
  \caption{(a)~Schematic of the ESN algorithms.\label{fig:fig1_intro}}
\end{figure}
  
In Fig.~\ref{fig:fig1_new} we plot the the response of the bubble to a positive square pulse [panel~(a.i)] and a negative square pulse [panel~(b.i)] (the response is given by the acoustic pressure scattered by the bubble according to Eq.~(\ref{eq:eq3})). The red curve denotes the response of the bubbles subjected to a step pressure change and the blue curve denotes the response of the bubble returned to its original equilibrium state. The pressure amplitude is $\alpha=0.2P_0$ in panel~(a.i) and $\alpha=-0.2P_0$ in panel (b.i). Panel~(c) shows the spectra of the bubble oscillations at $\alpha=0.2P_0$ (the top curves) and $\alpha=-0.2P_0$ (the bottom curves) alongside the spectra calculated at several values within the $-0.2P_0<\alpha<0.2P_0$ range.

The oscillations frequencies of the bubble are given by the frequencies of the main peaks in the spectra in panel~(c). In agreement with the analysis of the linearised KM equation (\ref{eq:eq1_linear}), the frequency of the bubble oscillating near its original equilibrium state (the blue curves) remains unchanged, but that of the bubble forced by the positive and negative step changes in the acoustic pressure (the red curves) increases and decreases, respectively. In panels~(a.ii) and (b.ii), we also plot the envelopes of the transient responses of the bubble and we observe changes in the relaxation time consistent with those predicted theoretically. In panel~(c) we observe that the numerical simulations predict the appearance of the second harmonic of the natural bubble oscillation frequency due to the nonlinearity (the third and fourth harmonics were also identified; however they are not shown on the figure). Respective changes in the relaxation time can also be noticed comparing, for example, the lineshapes of the fundamental and second-harmonic frequency peaks at $\alpha=\pm 0.2P_0$.
\begin{figure}[t]
  \centerline{\includegraphics[width=0.5\textwidth]{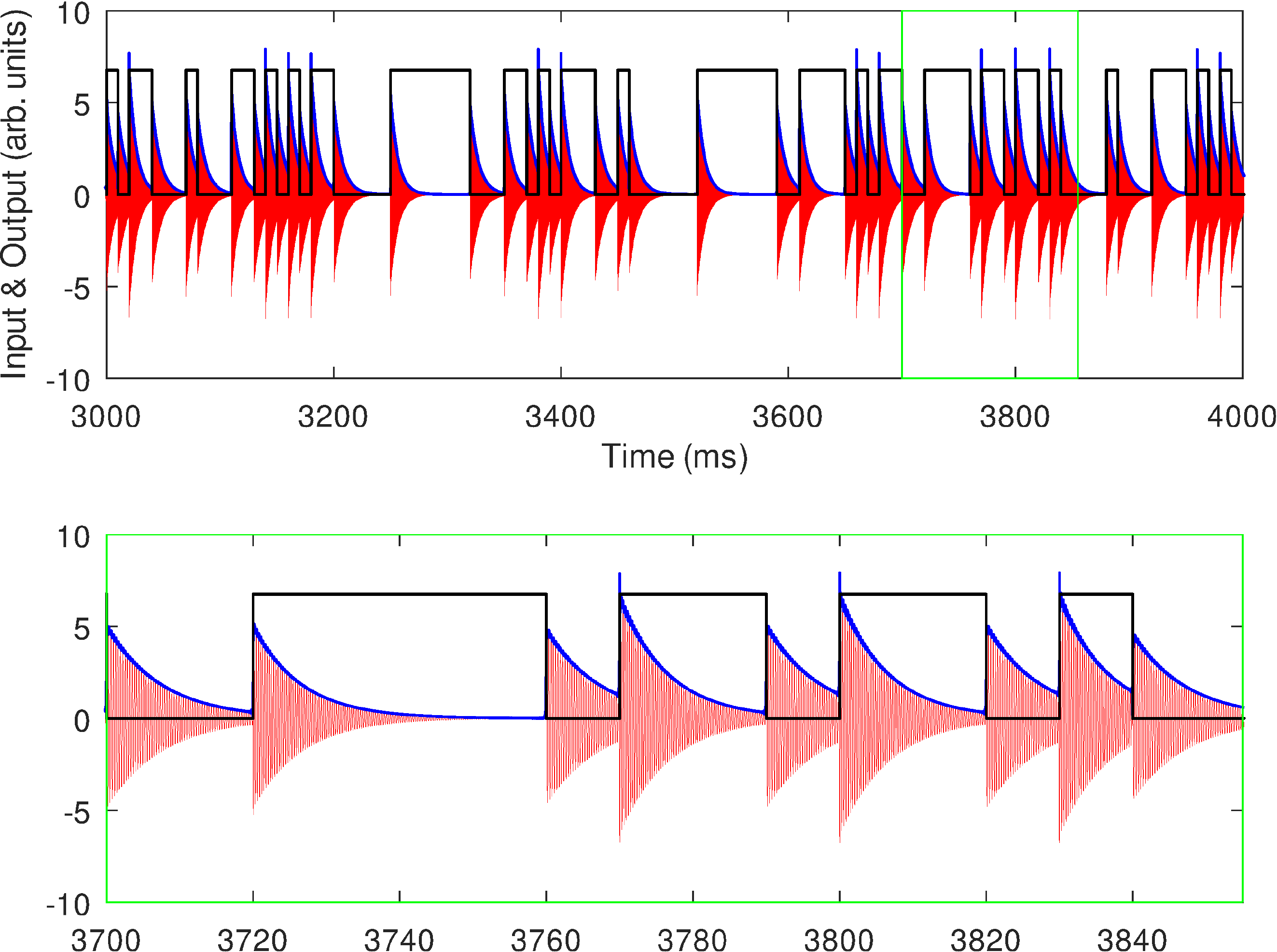}}
  \caption{Acoustic response of a single oscillating bubble (the red curve) to a sequence of square pressure pulses that encode a randomly generated binary time series (the black curve) discussed in the main text. The blue curve outlines the envelope of the bubble's response. In the top panel, the green rectangle highlights the time range of a magnified portion of the signals plotted in the bottom panel.\label{fig:fig1}}
\end{figure}

\section{Echo Sate Network\label{sec:ESN}}
To support our discussion of the ability of the bubble to operate as an artificial neuron, we use the standard Echo State Network (ESN) algorithm \cite{Jae05, Luk09}. We choose ESN because, alongside the relevant concept of Liquid State Machine (LSM) \cite{Maa02}, this type of neural network can be used to simulate some of the functions of a biological brain \cite{Ber04, Cuc22}.

In ESN, artificial neurons form an intricate network called the reservoir (Fig.~\ref{fig:fig1_intro}). The dynamics of the reservoir is governed by the nonlinear update equation
\begin{eqnarray}
  {\bf x}_{n} = (1-\alpha_{r}){\bf x}_{n-1}+
  \alpha_{r}\tanh({\bf W}^{in}{\bf u}_{n}+{\bf W}{\bf x}_{n-1})\,,
  \label{eq:RC1}
\end{eqnarray}
where $n$ is the index denoting equally-spaced discrete time instances $t_n$, ${\bf u}_n$ is the vector of $N_u$ input values, ${\bf x}_n$ is a vector of $N_x$ neural activations of the reservoir, the operator $\tanh(\cdot)$ is a sigmoid activation function of a neuron \cite{Luk09}, ${\bf W}^{in}$ is the input matrix containing $N_x \times N_u$ elements, ${\bf W}$ is the recurrent weight matrix consisting of $N_x \times N_x$ elements and $\alpha_{r} \in (0, 1]$ is a parameter that controls the reservoir's temporal dynamics \cite{Luk09}.

To train the reservoir, one typically calculates the output weights ${\bf W}^{out}$ by solving a system of linear equations ${\bf Y}^{target} = {\bf W}^{out}{\bf X}$, where ${\bf X}$ and ${\bf Y}^{target}$ are the state matrix and the target matrix that are constructed using, respectively, ${\bf x}_n$ and the vector of target outputs ${\bf y}_n^{~target}$ as columns for each discrete time instant $t_n$. The solution is often obtained in the form ${\bf W}^{out} = {\bf Y}^{target} {\bf X^\top} ({\bf X}{\bf X^\top} + \beta {\bf I})^{-1}$, where ${\bf I}$ is the identity matrix, $\beta \geq 0$ is a regularisation coefficient and ${\bf X^\top}$ is the transpose of ${\bf X}$ \cite{Luk09}. During the prediction stage, one solves Eq.~(\ref{eq:RC1}) for new input data ${\bf u}_n$ and computes the output vector ${\bf y}_n={\bf W}^{out}[1;{\bf u}_n;{\bf x}_n]$ using a constant bias and the concatenation $[{\bf u}_n;{\bf x}_n]$.
 
\section{Memory capacity of an oscillating bubble\label{sec:Memory}}
We demonstrate that a nonlinearly oscillating bubble retains memory of the past acoustic excitation inputs. We adopt the approach used to test the memory capacity of ESN \cite{Ber04, Jae05}. We excite oscillations of a single bubble with a randomly generated time series that contains binary elements that accept either '0' or '1' values. An individual '1' value is represented by a positive pressure square pulse of a unit length but a '0' value corresponds to a period of no acoustic excitation that also has the unit length duration. For example, an input data subset '10101' is represented by three square pulses of unit length separated by two periods of no excitation that also have a unit length duration. Accordingly, a sequence '111001111' corresponds to a triple square pulse separated from the following quadruple square pulse by two consecutive periods of no excitation. The actual unit length of an individual square pulse corresponding to a single '1' state is chosen empirically to be of order of the relaxation time $\tau_0$ of the bubble. This choice helps ensure that the response of the bubble to one pulse does not fully decay until the bubble is presented to the next pulse of the random time series, thereby enabling the bubble to remember the past inputs.

In line with the concept of physical reservoir computing, where the nonlinear dynamical behaviour of the artificial neural network is replaced by the response of a physical nonlinear-dynamical system \cite{Tan19, Nak20, Nak21, Cuc22}, we replace Eq.~(\ref{eq:RC1}) with Eq.~(\ref{eq:eq4}), which means that the artificial neural excitations ${\bf x}_n$ become a function of the acoustic response of the oscillating bubble. However, unlike Ref.~\cite{Mak21_ESN}, where ESN was implemented using the nonlinear dynamical response of a cluster of oscillating bubbles in water and where the individual bubbles of the cluster were operating as individual artificial neurons, in this work we employ a single oscillating bubble as a reservoir and we sample the bubble's temporal response to obtain 20 virtual artificial neurons that produce the values of ${\bf x}_n$ (for a relevant discussion see \cite{Wat20}).

We randomly generate a time series that consists of 2000 binary '0' and '1' values. We encode each binary value as square pulses and no excitation gaps between them, respectively. The amplitude of the square pulses is $\alpha=0.1P_0$. In the framework of ESN, the so-encoded signal plays the role of the input vector ${\bf u}_n$.

We calculate the response of the bubble to the whole input signal and then we sample the bubble's response to obtain the neural activation ${\bf x}_n$ corresponding to 20 virtual neurons. Afterwards we use the first half of the input and the first half of the respective bubble's response as the training data, and we calculate ${\bf W}^{out}$ using the procedure described in Sect~\ref{sec:ESN}. The second half portions of these data sets are used to conduct the memory test.
\begin{figure}[t]
  \centerline{\includegraphics[width=0.45\textwidth]{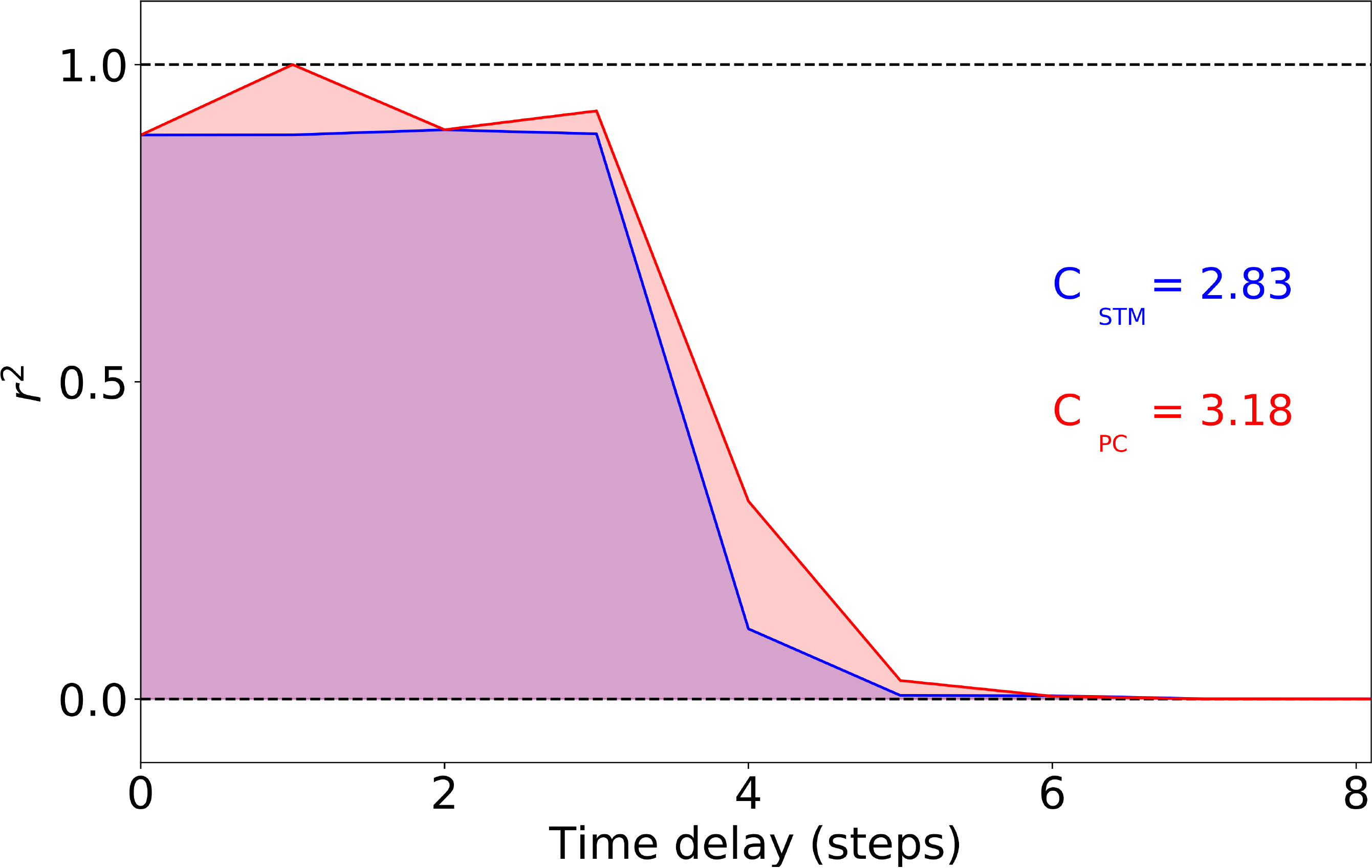}}
  \caption{STM and PC task memory capacity of an oscillating bubble expressed as the square of the correlation coefficient $r^2$ calculated for different discrete time delays. \label{fig:fig2}}
\end{figure}

Figure~\ref{fig:fig1} shows a typical response of the bubble to the input time series represented by the square pulses and gaps between them. Each square pulse followed by a no excitation period causes a step change in the pressure of the driving acoustic wave, thereby triggering two bursts of bubble oscillation: one at the beginning of the pulse and another one at the end of it. As a result, the oscillations caused by one pulse do not fully decay until the bubble is excited by the following pulse, which means that the state of the system depends on the past inputs. In turn, this indicates that the system possesses certain memory \cite{Jae05, Wat20}, which is demonstrated in continuation.

We employ two standard approaches to calculate the memory capacity of an oscillating bubble. The first of them is the $k$-delay task applied to a random binary time series, where we calculate the correlation coefficient $r({\bf y}^{target}_{n-k}, {\bf y}_{k,n})$ between a delayed target signal ${\bf y}^{target}_{n-k}$ and the predicted output ${\bf y}_{k,n}$ produced by the oscillating bubble trained using the training data ${\bf u}_{k,n}$ with a discrete time delay $k$ \cite{Jae05}. The function $r^{2}({\bf y}^{target}_{n-k}, {\bf y}_{k,n})$ can accept any value from 0 to 1, being 0 an indicator of the full loss of correlation and 1 corresponding to a full correlation between the correct target data and the prediction made by the bubble. Calculating $r^2$ as a function of the discrete delay time $k$, we determine the so-called short-term (also known as fading) memory capacity (STM) \cite{Jae05, Wat20} as
\begin{eqnarray}
  C_{STM} = \sum_{k=1}^{k_{max}} r^{2}({\bf y}^{target}_{n-k}, {\bf y}_{k,n})\,,
  \label{eq:RC5}
\end{eqnarray}
being $k_{max}$ the maximum delay considered in the analysis.

Then, we task the oscillating bubble with a parity check test \cite{Ber04}, where the bubble predicts a time series defined as $PARITY({\bf u}_{n-k}, {\bf u}_{n-k-1}, {\bf u}_{n-k-2})$ for increasing delays $k$. Based on the original randomly generated binary time series, the so-generated new time series is a binary sequence itself. The parity function is not linearly separable and its prediction requires the reservoir to have memory. Similarly to Eq.~(\ref{eq:RC5}), we calculate the parity check memory capacity $C_{PC}$ by summation of the parity check values up to the maximum delay $k_{max}$.

The calculated memory capacities are presented in Fig.~\ref{fig:fig2}, where we plot $r^2$ as a function of the discrete delay time. We obtain $C_{STM}=2.83$\,bits and
$C_{PC}=3.18$\,bits. These values are similar to those obtained for physical reservoir systems of other nature \cite{Wat20} and they confirm that an oscillating bubble has a decent memory capacity. We also note in Fig.~\ref{fig:fig2} at the zero delay both STM and PC tests produce the memory of approximately $0.9$\,bit instead of the theoretically expected 1\,bit corresponding to 100\% correlation. In the standard ESN approach, this situation originates from long-term reverberations in the reservoir \cite{Kir91} that deteriorate the recall of immediately past inputs \cite{Jae05}.
        
\end{document}